\documentclass[preprint,preprintnumbers,amsmath,amssymb]{revtex4}

\usepackage{graphicx}% Include figure files
\usepackage{dcolumn}% Align table columns on decimal point
\usepackage{bm}% bold math
\usepackage{braket}
\begin{document}

%\preprint{APS/123-QED}

\title{Monte Carlo simulation of a statistical mechanical model of multiple protein sequence alignment}% Force line breaks with \\

\author{Akira R. Kinjo}
\email{akinjo@protein.osaka-u.ac.jp}
\affiliation{Institute for Protein Research, Osaka University\\
  3-2 Yamadaoka, Suita, Osaka, 565-0871, Japan.}%Lines break automatically or can be forced with \\

\date{\today}% It is always \today, today,

\begin{abstract}
A grand canonical Monte Carlo (MC) algorithm is presented for studying the lattice gas model (LGM) of multiple protein sequence alignment, which coherently combines long-range interactions and variable-length insertions. MC simulations are used for both parameter optimization of the model and production runs to explore the sequence subspace around a given protein family. In this Note, I describe the details of the MC algorithm as well as some preliminary results of MC simulations with various temperatures and chemical potentials, and compare them with the mean-field approximation. The existence of a two-state transition in the sequence space is suggested for the SH3 domain family, and inappropriateness of the mean-field approximation for the LGM is demonstrated.
\end{abstract}

%\pacs{}% PACS, the Physics and Astronomy
\maketitle

\section{Introduction}
Thanks to the massive genome sequencing, we have a great number of known amino acid sequences at our hands. Exploiting the wealth of sequence data, recent advances in biological sequence analysis made it possible to reliably extract the ``direct'' couplings between residues that are separated along the sequence \cite{MorcosETAL2011,JonesETAL2012,TaylorETAL2012,Miyazawa2013,Levy2017}, and thereby to accurately predict three-dimensional (3D) structures \cite{MarksETAL2011,Ovchinnikov2017} as well as mutation effects \cite{HopfETAL2017}.

Based on these developments, I have previously proposed a lattice gas model (LGM) of multiple protein sequence alignment (MSA) which incorporates direct couplings and variable-length insertions in a coherent manner \cite{Kinjo2016}. In that work, I tried to use a mean-field approximation for treating the long-range direct couplings so that the partition function can be computed efficiently. However, it was found that the quasi-1-dimensional model structure was somehow incompatible with long-range interactions so that the original mean-field approximation without the diagonal terms of the direct coupling matrix failed to converge to correct solutions. Thus, I resorted to the Gaussian approximation by including the diagonal terms of the ``long-range'' interaction matrix. The Gaussian approximation makes the system essentially harmonic, and hence, by construction, it does not exhibit some interesting phenomena such as phase transitions. In a preliminary study, I also tested the pseudolikelihood method \cite{BalakrishnanETAL2011,EkebergETAL2013} for obtaining the direct couplings, but again, failed to obtain stable solutions (data not shown).
Therefore, for the LGM of protein families, it appears necessary to drop any approximations (at least those known to the author). In principle, this can be done by performing Monte Carlo (MC) simulations, which was indeed the original approach to the problem \cite{LapedesETAL1999}, and more recently employed by Sutto et al.\cite{SuttoETAL2015}. In these studies, only alignments with fixed lengths were treated so that the standard canonical Metropolis sampling was sufficient. In the LGM, however, the alignment length is variable, which necessitates some special treatment. In short, it requires a special kind of grand canonical MC algorithm which I believe is worth sharing in this Note. Along the way, I reformulate the LGM from a different perspective, which may help better understand the physical meaning of the model.

\section{Materials and Method}
\subsection{Model}
\begin{figure*}
  \begin{center}
    \includegraphics[width=\textwidth]{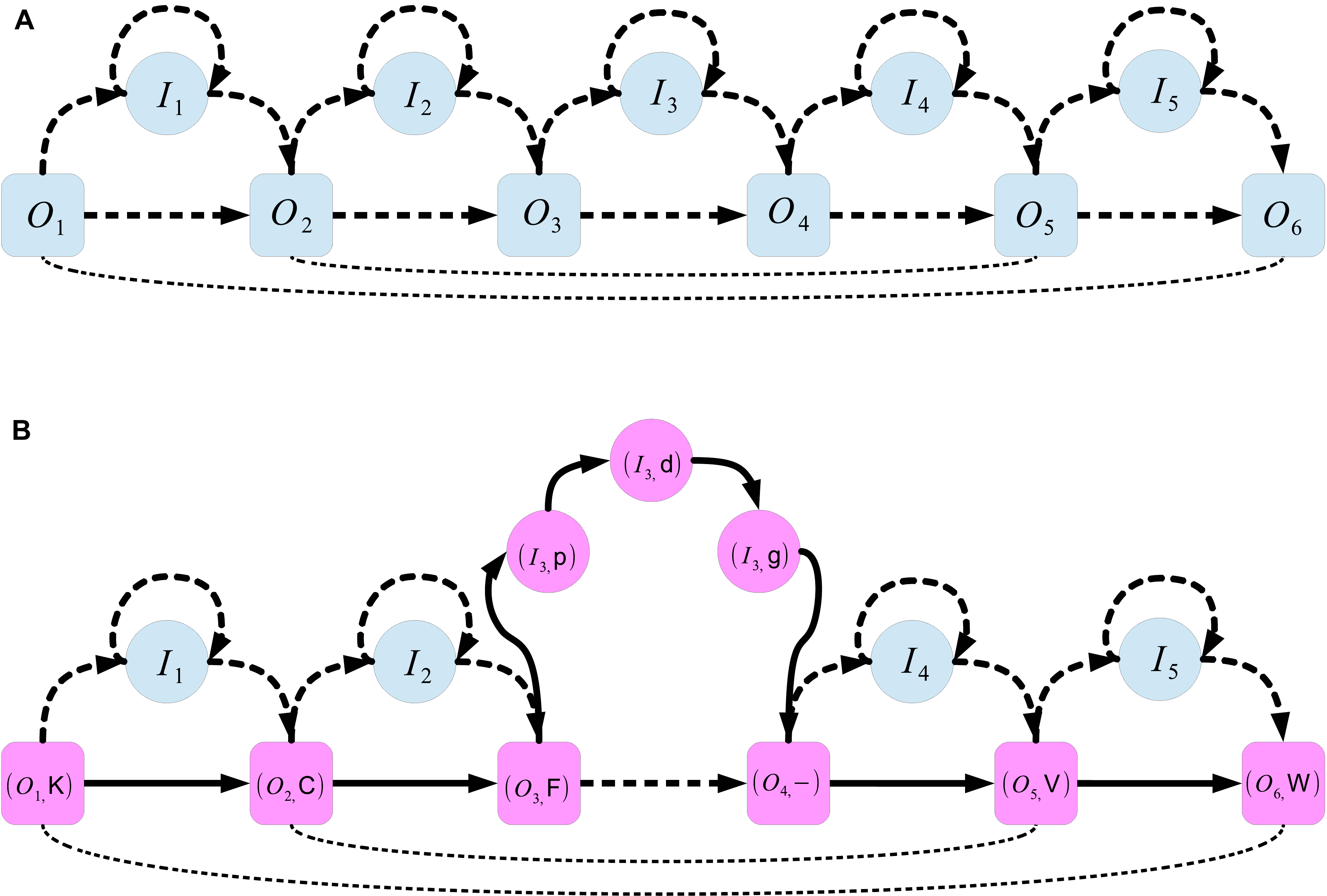}
  \end{center}
  \caption{\label{fig:model} An example of the model structure.
    (A) A model with $N = 6$. The potentially bonded pairs are connected via dashed arrows, and interacting pairs ($\mathcal{T} = \{(O_1,O_6),(O_2,O_5)\}$) are connected via dashed lines.
    (B) One possible alignment between the model in (A) and an amino acid sequence \texttt{KCFPDGVW} is represented as
$\mathbf{X} = X_{1}\cdots X_{9} =$
$(O_1,\mathtt{K})$%
$(O_2,\mathtt{C})$%
$(O_3,\mathtt{F})$%
$(I_3,\mathtt{p})$%
$(I_3,\mathtt{d})$%
$(I_3,\mathtt{g})$%
$(O_4,\mathtt{-})$%
$(O_5,\mathtt{V})$%
    $(O_6,\mathtt{W})$.
Here we adopt the conventions that residues aligned with insert sites are written in the lower case and bonded pairs are connected via solid arrows.  Note the insert site $I_3$ appears more than once in this particular alignment.
  }
\end{figure*}
The basic structure of the LGM was presented in a previous paper \cite{Kinjo2016}. Here, I reformulate the model more formally with a few modifications.
An LGM $\mathcal{M}$ is defined as a tuple of sets $\mathcal{M} = \{\mathcal{S}, \mathcal{A}, \mathcal{T}\}$. The set $\mathcal{S}$ is a set of $2N-1$ \emph{model sites}, which is a disjoint union of the set of $N$ \emph{core sites} $\mathcal{O} = \{O_1,\cdots,O_N\}$ and the set of $N-1$ \emph{insert sites} $\mathcal{I} = \{I_1,\cdots, I_{N-1}\}$: $\mathcal{S} = \mathcal{O}\cup\mathcal{I}$.
The number of core sites, $N$, is called the \emph{length} of the model.
The core sites represent those sequence positions that are present in a majority of sequences of the protein family of interest, and the insert sites represent the other positions. In this study, I define the core and insert sites as the positions of the ``match'' and ``insert'' states, respectively, of the profile hidden Markov model (HMM) of the corresponding Pfam \cite{Pfam} family.
The set $\mathcal{S}$ is partially ordered (denoted by ``$\to$'') such that $O_i \to O_{i+1}$, $O_i \to I_i$, $I_i \to I_i$ and $I_i \to O_{i+1}$ for $i = 1,\cdots,N-1$, and these ordered pairs are said to be \emph{potentially bonded} (the dashed arrows in Fig. \ref{fig:model}A).
Those pairs of model sites that are not potentially bonded are called \emph{non-bonded}. The set $\mathcal{A}$ is a set of sets of allowed amino acid residue types for each model sites: $\mathcal{A} = \{\mathcal{A}_S | S \in \mathcal{S}\}$ where $\mathcal{A}_{O_i} (i = 1,\cdots, N)$ contains the 20 standard residues types and a symbol for the ``delete'' (``-'') and $\mathcal{A}_{I_i} (i = 1,\cdots, N-1)$ contains the 20 standard residue types. The set $\mathcal{T}$ is a set of \emph{interacting non-bonded pairs} of core sites. To define it precisely, we first define the (non-redundant) set of all pairs of core sites: ${\mathcal{T}^{*}} = \{(O_i,O_j) | O_i, O_j \in \mathcal{O}, i < j\}$. Then, $\mathcal{T}$ is defined as a subset: $\mathcal{T} \subset \mathcal{T}^{*}$. This subset is determined based on structural information (see below). A concrete example of a model with $N=6$ is shown in Fig. \ref{fig:model}A.

Using this model, we can represent an alignment between an arbitrary amino acid sequence and the model. 
Let $\mathbf{a} = a_1\cdots a_L$ be an amino acid sequence of $L$ residues. An alignment between the sequence $\mathbf{a}$ and the model $\mathcal{M}$ is represented as a sequence of pairs of a model site (core or insert) and a residue: $\mathbf{X} = X_1\cdots X_{L_{\mathbf{X}}}$ where
$L_{\mathbf{X}}$ is the length of the alignment and each $X_i$ is an \emph{aligned pair} such as $(S,a)$ with $S\in \mathcal{S}$ and $a \in \mathcal{A}_S$,
$a = a_j$ for some $j \in \{1,\cdots, L\}$ or $a = \text{``-''}$ (delete, only if $S$ is a core site). For $\mathbf{X}$ to be a proper alignment, there are two requirements. First, for any two consecutive aligned pairs $X_kX_{k+1}=(S,a)(S',a')$, the two model sites must be potentially bonded ($S \to S'$). Note in particular that the same insert site can appear arbitrarily many times in an alignment due to the order $I_i \to I_i$. The two model sites that appear in two consecutive aligned pairs are called \emph{bonded} (note the absence of the adverb ``potentially'').
Second, for any two aligned pairs $X_k=(S,a_i)$ and $X_l=(S',a_j)$ in $\mathbf{X}$ that are aligned with proper amino acid residues $a_i$ and $a_j$ in the sequence $\mathbf{a}$, if $k<l$ then $i < j$.
A concrete example of an alignment is shown in Fig. \ref{fig:model}B. 

In the present study, I have made a few modifications (Fig. \ref{fig:model}) to the previous model \cite{Kinjo2016}. The first is that the alignment is now global with respect to the model but local with respect to the sequence (the previous model was global with respect to both the model and sequence). This means that the entire region of a model is always aligned with an amino acid sequence while only a part of the amino acid sequence may be aligned with the model (by ignoring possible flanking residues on both the N- and C-termini of the amino acid sequence).

Another major modification is the limited number of interacting non-bonded pairs (previously, all the core sites were interacting with each other: $\mathcal{T} = \mathcal{T}^{*}$).
Between core sites more than 5 residues apart along the sequence, there may be interactions defined based on a representative native structure of the family. Two sites are defined to be interacting if the residues aligned to those sites are in contact in the corresponding (representative) native structure. Two residues are defined to be in contact if any non-hydrogen atoms in those residues are within 5\AA. Interactions are defined only between core sites for simplicity.

Based on this representation of the alignment, the energy function of the alignment $\mathbf{X}$ is given as
\begin{equation}
  \label{eq:energy0}
  E(\mathbf{X}) =
  -\sum_{k=1}^{L_{\mathbf{X}}-1}J(X_k,X_{k+1})
  -\sum_{(s(X_k),s(X_l)) \in \mathcal{T}}K(X_k,X_l)
  -\sum_{k=1}^{L_{\mathbf{X}}}\mu(X_k)
\end{equation}
where $J$ and $K$ are, respectively, short-range and long-range interaction energy parameters to be determined from the given (observed) MSA and $\mu$ is the chemical potential, and $\mathcal{T}$ indicates the set of all interacting pairs of core sites and $s$ is a function to extract the model site from an aligned pair (i.e., $s(X) = S$ for $X = (S,a), S\in \mathcal{S}$). Thus, only $J$ and $K$ parameters constitute the intrinsic energy, and $\mu$'s are provided as external variables to control (perturb) the system.

We assume that the probability $P(\mathbf{X})$ of obtaining an alignment $\mathbf{X}$ is given by the Boltzmann distribution:
\begin{equation}
  \label{eq:boltzmann}
  P(\mathbf{X}) = \frac{\exp[-E(\mathbf{X})/T]}{\Xi[T]}
\end{equation}
where $T$ is the ``temperature'' and $\Xi[T]$ is the partition function 
\begin{equation}
\Xi[T] = \sum_{\mathbf{X}}\exp[-E(\mathbf{X})/T]\label{eq:partition-function}.
\end{equation}
Here, the summation is over all possible alignments ($\mathbf{X}$) with the model and all possible sequences. Since the alignment length can vary, this ensemble is considered to be a grand canonical ensemble. The grand potential is given by
\begin{equation}
  \label{eq:grandpot}
  \Omega[T] = -T\log\Xi[T].
\end{equation}

Given the parameters $J$, $K$ and $\mu$, we can sample sequences according to the probability distribution Eq. (\ref{eq:boltzmann}) by running Monte Carlo (MC) simulations.

The \emph{standard condition} is defined to be the system with $\mu(X_k) = 0$ for all $X_k$, and the \emph{natural condition} to be the standard condition with $T = 1$. 

The parameters $J$ and $K$ are determined iteratively so that the average residue pair counts for bonded and interacting non-bonded pairs over the samples produced by MC simulations under the natural condition match those observed in the given MSA of the family. However, this procedure is not straightforward since the length of alignment $\mathbf{X}$ is variable. The complication is due to the indexing scheme of aligned pairs where the meaning of each index, say ``$k$'' of $X_k$, is different for different alignments. Therefore, I reformulate the energy function by introducing some stochastic variables, single-site counts and bonded pair counts, based on model sites which are fixed for any alignments. In fact, this was the original formulation of the LGM in the previous paper \cite{Kinjo2016}.

For each model site and site pairs, we define the stochastic variables as functions of alignment $\mathbf{X} = X_1\cdots X_{L_{\mathbf{X}}}$.
To do so, we first define the set $\mathcal{Y}$ of all possible pairs of model sites and amino acid residues as $\mathcal{Y} = \{(S,a) | S \in \mathcal{S}, a \in \mathcal{A}_{S}\}$. Note that, while the set of all possible alignments $\{\mathbf{X}\}$ is an infinite set, the set $\mathcal{Y}$ is finite (namely, $|\mathcal{Y}| = 21N + 20(N-1)$).
Now, the single-site count for the pair $Y \in \mathcal{Y}$ is defined as
\begin{equation}
  \label{eq:sscount}
  n_{Y}(\mathbf{X}) = \sum_{k=1}^{L_{\mathbf{X}}}\delta_{Y,X_k}
\end{equation}
where $\delta_{Y,X_k}$ indicates Kronecker's delta (i.e., $\delta_{Y,X_k} = 1$ if $Y = X_k$ and $\delta_{Y,X_k} = 0$ otherwise). Note that $n_{(O_i,a)}(\mathbf{X})$ can be either 0 or 1 for any core site $O_i$ whereas $n_{(I_i,a)}(\mathbf{X})$ for any insert site $I_i$ may have any values from 0 to infinity.
Next, the bonded pair count for $Y,Y' \in \mathcal{Y}$ is given as
\begin{equation}
  \label{eq:bpcount}
n_{Y,Y'}^b(\mathbf{X}) = \sum_{k=1}^{L_\mathbf{X}-1} \delta_{Y,X_{k}} \delta_{Y',X_{k+1}}.
\end{equation}
The bonded pair counts are defined only for potentially bonded pairs of model sites (those pairs of sites connected via dotted arrows in Fig. \ref{fig:model}A). 
However, two neighboring model sites may not always be aligned with two consecutive residues in the amino acid sequence. For example, in the alignment $\mathbf{X}$ shown in Fig. \ref{fig:model}B, we have the bonded pair count $n_{(O_3,\mathtt{F}),(O_4,\mathtt{-})}^b(\mathbf{X}) = 0$ because the pair of aligned sites $(O_3,\mathtt{F})$ and $(O_4, \mathtt{-})$ are not consecutive in the alignment $\mathbf{X}$ although the model sites $O_3$ and $O_4$ are potentially bonded in the model. Nevertheless, we have $n_{(O_3,\mathtt{F})}(\mathbf{X})n_{(O_4,\mathtt{-})}(\mathbf{X}) = 1$ because $n_{(O_3,\mathtt{F})}(\mathbf{X})=1$ and $n_{(O_4,\mathtt{-})}(\mathbf{X}) = 1$.
Therefore, a bonded pair count cannot be reduced to a product of two single-site counts. In this manner, bonded pair counts account for the chain structure of polypeptide sequences, which should not be confused with the (quasi-)one-dimensional lattice structure of the model.

From the definitions of the single-site and bonded pair counts, we have the following relations \cite{Kinjo2016}. First, the single-site counts for each core site are normalized:
\begin{eqnarray}
  \sum_{a\in \mathcal{A}_{O_i}}n_{(O_i,a)}(\mathbf{X}) &=& 1. \label{eq:rel1}
\end{eqnarray}
Second, each single-site count is completely determined by bonded pair counts:
\begin{eqnarray}
    \sum_{b\in\mathcal{A}_{O_{i+1}}}n_{(S,a),(O_{i+1},b)}^{b}(\mathbf{X})
  + \sum_{b\in\mathcal{A}_{I_{i}}}n_{(S,a),(I_{i},b)}^{b}(\mathbf{X}) & = & n_{(S,a)}(\mathbf{X}),\label{eq:relb1}\\
  \sum_{a\in\mathcal{A}_{O_{i}}}n_{(O_{i},a),(S',b)}^{b}(\mathbf{X})
  + \sum_{a\in\mathcal{A}_{I_{i}}}n_{(I_{i},a),(S',b)}^{b}(\mathbf{X}) & = & n_{(S',b)}(\mathbf{X}).\label{eq:relb2}
\end{eqnarray}
Finally, it follows from the relation Eq. (\ref{eq:rel1}) that pair counts for non-bonded pairs also determine the single-site counts:
\begin{eqnarray}
  \sum_{b\in\mathcal{A}_{O_j}}n_{(O_i,a)}(\mathbf{X})n_{(O_j,b)}(\mathbf{X}) &=& n_{(O_i,a)}(\mathbf{X}),  \label{eq:relnb1}\\
  \sum_{a\in\mathcal{A}_{O_i}}n_{(O_i,a)}(\mathbf{X})n_{(O_j,b)}(\mathbf{X}) &=& n_{(O_j,b)}(\mathbf{X}).  \label{eq:relnb2}
\end{eqnarray}
These relations explain why chemical potentials $\mu$'s are not necessary as a part of the intrinsic energy parameters (Eq. \ref{eq:energy0}).
They also indicate that not all the variables are independent, which in turn indicates there is gauge freedom in the energy parameters $J$ and $K$. In this study, I (partially) fixed the gauge so that $J((O_i,\mathtt{-}),(O_{i+1},\mathtt{-})) = 0$ and $K((O_i,\mathtt{-}),(O_{i+1},\mathtt{-})) = 0$, that is, the interactions between the ``delete'' residues (``$\mathtt{-}$'') were defined to be zero.

Using the single-site counts and bonded pair counts, we can rewrite the energy function as
\begin{eqnarray}
  E(\mathbf{X}) &=&  -\sum_{Y,Y'}^{b.p.}J(Y,Y')n_{Y,Y'}^b(\mathbf{X}) \nonumber\\
  & &  -\sum_{Y,Y'}^{n.b.p.}K(Y,Y')n_{Y}(\mathbf{X})n_{Y'}(\mathbf{X})\nonumber\\
  & &  -\sum_{Y\in\cal{Y}}\mu(Y)n_{Y}(\mathbf{X}) \label{eq:energy}
\end{eqnarray}
where the summations $\sum_{Y,Y'}^{b.p.}$ and $\sum_{Y,Y'}^{n.b.p.}$ are taken over all the possible pairs of aligned pairs over potentially bonded pairs and interacting non-bonded pairs in the model sites, respectively.

\subsection{Monte Carlo algorithm}
\begin{figure}
  \includegraphics[width=0.5\textwidth]{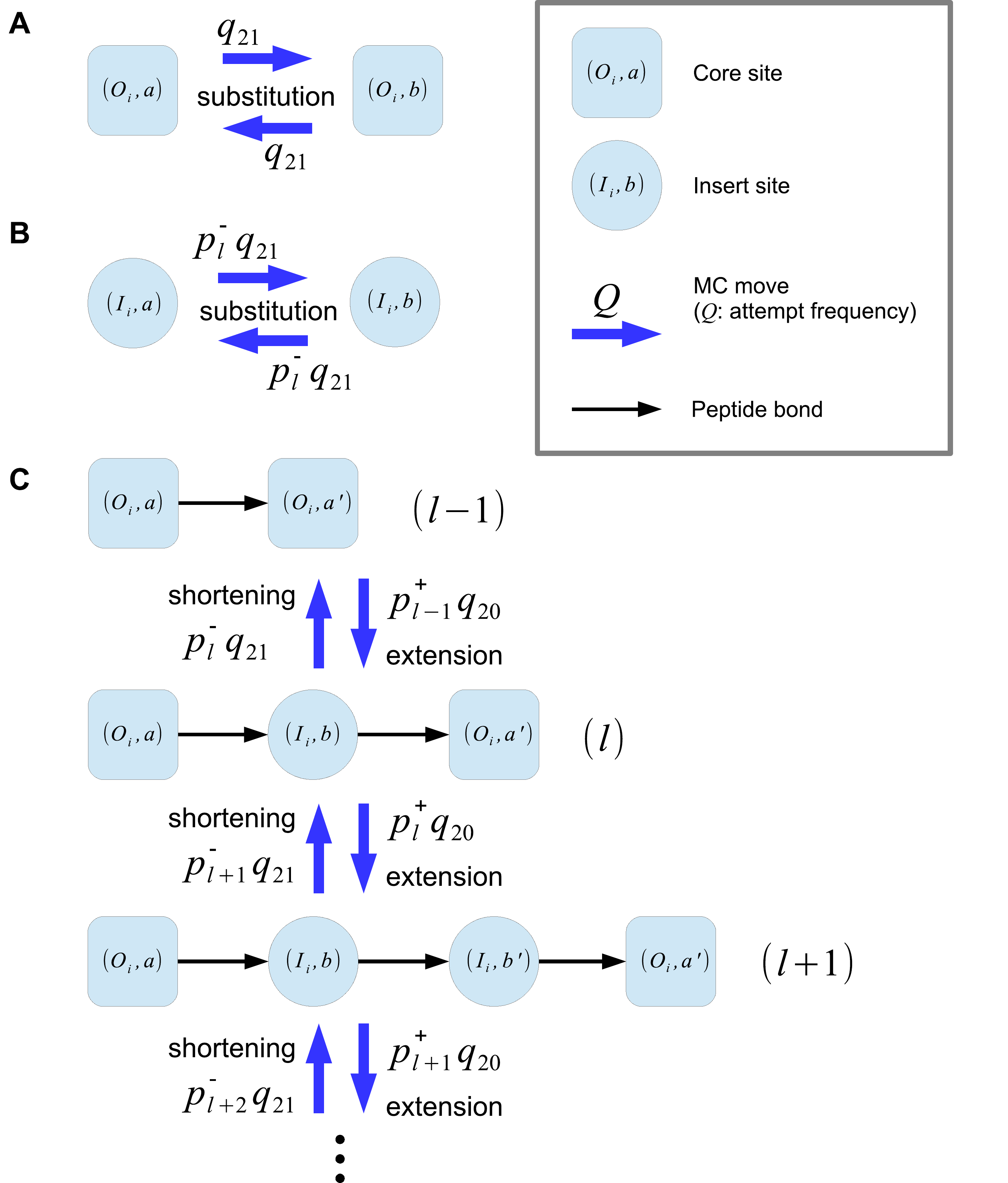}
  \caption{\label{fig:mcmove}MC moves. For an LGM model of length $N$, each core or insert site is randomly selected with probability $N/(2N-1)$ or $(N-1)/(2N-1)$, respectively. (A) At a core site, a substituting residue is chosen with probability $q_{21} = 1/21$, which is accepted with the probability given in Eq. \ref{eq:metropolis}.
    (B) For a substitution at an insert site aligned with $l$ residues, a tentative substituting residue at one of the $l$ residue positions is randomly chosen with the probability $p_{l}^{-}q_{21}$ where $p_{l}^{-} = l/(2l+1)$, which is accepted with probability given in Eq. \ref{eq:metropolis}.
    (C) An insert may be extended by adding a residue or be shortened by deleting a residue. The extension of an insert with $l$ residues is attempted with probability $p_{l}^{+}q_{20}$ where $p_{l}^{+} = (l+1)/(2l+1)$ and $q_{20} = 1/20$, which is accepted with the probability given in Eq. (\ref{eq:pextend}).
  The shortening of an insert with $l$ residues is attempted with probability $p_{l}^{-}q_{21}$, which is accepted with the probability given in Eq. (\ref{eq:pshorten}).}
\end{figure}

I first describe the Metropolis-Hastings algorithm \cite{Iba2005} for sampling alignments $\{\mathbf{X}\}$. Since alignment length $L_{\mathbf{X}}$ is variable depending on insertions, some care is necessary to ensure the detailed balance.

For a core site, we randomly select a residue $b$ out of 21 possible residue types, substitute it with the existing residue $a$ (Fig. \ref{fig:mcmove}A). Let the energies of the sequences before and after the substitution be $E_a$ and $E_{b}$, respectively, the substitution is accepted with the probability
\begin{equation}
\min\{1, \exp[-(E_{b}-E_a)/T]\}.  \label{eq:metropolis}
\end{equation}

For an insert site, there are three possible moves: substituting a residue (Fig. \ref{fig:mcmove}B), extending the insert by adding a new residue or shortening the insertion by deleting a residue (Fig. \ref{fig:mcmove}C). Suppose an insert consists of $l$ residues ($l = 0,1,\cdots$). Then there are $l+1$ possible positions for a new residue (one of 20 types is chosen with probability $q_{20} = 1/20$) to be inserted between two consecutive residues, and $l$ possible positions (residues) for an existing residue to be deleted or substituted (deletion or a substituting residue is chosen with probability $q_{21}=1/21$). Accordingly, an extension of the insert is attempted with probability (attempt frequency)
$p_{l}^{+}q_{20}$ where $p_{l}^{+} = (l+1)/(2l+1)$,
and a deletion or substitution is attempted with probability $p_{l}^{-}q_{21}$ where $p_{l}^{-} = l/(2l+1) = 1 - p_{l}^{+}$. This choice of attempt frequencies automatically excludes the possibility of attempting a deletion or substitution when $l = 0$.

For an extension of an insert, one of the $l+1$ positions and one of the 20 residue types are chosen randomly, then it is inserted to the selected position. Let the energies before and after the extension be $E_l$ and $E_{l+1}$, respectively. Then the extension is accepted with the probability
\begin{equation}
\min\left\{1, \frac{p_{l+1}^{-}q_{21}}{p_{l}^{+}q_{20}}\exp[-(E_{l+1}-E_l)/T]\right\}  \label{eq:pextend}
\end{equation}
where $p_{l+1}^{-} = (l+1)/(2l+3)$ (the probability of attempting substitution or deletion for an insert of length $l+1$), $q_{21} = 1/21$ (uniform distribution for selecting a substituting residue or deletion) and $q_{20} = 1/20$ (uniform distribution for selecting an inserted residue). The pre-exponential factor ensures the detailed balance condition \cite{Iba2005}.

For a substitution or deletion of an insert, randomly pick one of the $l$ positions in the insert. A deletion is attempted with probability $q_{21} = 1/21$. Let the energy after the attempted deletion be $E_{l-1}$. Then the deletion is accepted with probability
\begin{equation}
\min\left\{1, \frac{p_{l-1}^{+}q_{20}}{p_{l}^{-}q_{21}}\exp[-(E_{l-1} - E_l)/T]\right\}  \label{eq:pshorten}
\end{equation}
where $p_{l-1}^{+} = l/(2l-1)$ (the probability of attempting the extension of an insert of $l-1$ residues) and $p_{l}^{-} = 1 - p_{l}^{+}$. For a substitution, randomly pick a substituting residue with the uniform distribution ($q_{20} = 1/20$), then apply the usual Metropolis criteria (c.f., Eq. \ref{eq:metropolis}).

For substitutions at core and insert sites, Gibbs sampling was also employed during parameter optimization. In these cases, substituting residues are selected according to the probability proportional to $\exp[-E(a)/T]$ where $E(a)$ is the energy of the sequence with the attempted substitution with residue type $a$ (either in a core or insert site).

For a model of length $N$, one sweep consists of $N$ and $N-1$ moves for randomly chosen core and insert sites, respectively.

\subsection{Parameter optimization}
Let $\braket{Q}_{\mathrm{obs}}$ denote the average value of the variable $Q(\mathbf{X})$ over a given MSA. If the MSA consists of $M$ aligned sequences ($\mathbf{X}^1,\cdots,\mathbf{X}^{M}$), we compute the observed averages $\braket{n_{Y}}_{\mathrm{obs}}$, $\braket{n_{YY'}^b}_{\mathrm{obs}}$ and $\braket{n_{Y}n_{Y'}}_{\mathrm{obs}}$   as follow:
\begin{eqnarray}
  \braket{n_Y}_{\mathrm{obs}} &=& \frac{1}{\gamma+1}\left[\frac{\gamma}{q_Y}+\sum_{t=1}^M w_t{n_{Y}(\mathbf{X}^t)}\right],\\
  \braket{n_{YY'}^{b}}_{\mathrm{obs}} &=& \frac{1}{\gamma+1}\left[\frac{\gamma}{2q_Yq_{Y'}}+\sum_{t=1}^M w_t{n_{YY'}^{b}(\mathbf{X}^t)}\right],\\
  \braket{n_Yn_{Y'}}_{\mathrm{obs}} &=& \frac{1}{\gamma+1}\left[\frac{\gamma}{q_Yq_{Y'}}+\sum_{t=1}^M w_t{n_{Y}(\mathbf{X}^t)n_{Y'}(\mathbf{X}^t)}\right]
\end{eqnarray}
where $q_Y$ (or $q_{Y'}$) is 21 or 20 if the model site of the the aligned pair $Y$ (or $Y'$) is a core site or an insert site, respectively, and $w_t$ is the position-based weight of the sequence $t$ \cite{HenikoffANDHenikoff1994}. Only the core sites were used for computing the weights and the weights are normalized (i.e., $\sum_{t=1}^{M} w_t = 1$). These average counts are referred to as \emph{number densities} in the following.
The value of the pseudocount $\gamma$ was set to a relatively small value ($\gamma = 0.1$) in order to keep the expected sequence length with the pseudocounts closer to the value without the pseudocounts. 

We define $\braket{Q}_{\mathrm{sim}}$ as the average of the variable $Q(\mathbf{X})$ over a set of samples obtained from MC simulations (at a constant temperature). It is given as a simple average:
\begin{equation}
  \braket{Q}_{\mathrm{sim}} = \frac{1}{M}\sum_{t=1}^{M}Q(\mathbf{X}^{t}).
\end{equation}
for a set of $M$ simulated samples.

Given the estimates for observed and simulated number densities, we can optimized the energy parameters $J$ and $K$ (Eqs. \ref{eq:energy0} and \ref{eq:energy}). Under the natural condition ($\mu(Y) = 0$ for all $Y$ and $T=1$), the average energy of the observed MSA is given as
\begin{equation}
  \braket{E}_{\mathrm{obs}} =
    -{\sum_{(Y,Y')}^{\mathrm{b.p.}}}J(Y,Y')\braket{n_{Y,Y'}^b}_{\mathrm{obs}}
 -{\sum_{(Y,Y')}^{\mathrm{n.b.p.}}}K(Y,Y')\braket{n_{Y}n_{Y'}}_{\mathrm{obs}}.
\end{equation}
The parameter optimization is done by maximizing $\braket{E}_{\mathrm{obs}} - \Omega[T]$ under the natural condition ($T = 1$ in particular), or:
\begin{equation}
  \label{eq:free-energy}
  F[T] = -\max_{J,K}\left(\braket{E}_{\mathrm{obs}} - \Omega[T]\right)
\end{equation}
where $\Omega[T]$ is the grand potential defined in Eq. (\ref{eq:grandpot}).
This may be regarded as a Legendre transform from the grand canonical ensemble
with grand potential $\Omega[T]$ determined by $J(Y,Y')$ and $K(Y,Y')$ to the canonical ensemble with free energy $F[T]$ determined by the respective conjugate variables $\braket{n_{YY'}^b}_{\mathrm{obs}}$ and $\braket{n_Yn_{Y'}}_{\mathrm{obs}}$. The optimization of the parameters as given in this equation is equivalent to the principle of maximum entropy as used by others \cite{MorcosETAL2011} and in the previous paper \cite{Kinjo2016}.
Let $P$ denote either $J(Y,Y')$ or $K(Y,Y')$, and $Q(\mathbf{X})$ denote,
respectively, $n_{YY'}^b(\mathbf{X})$ or ${n_Y(\mathbf{X})n_{Y'}(\mathbf{X})}$.
We have the following relations:
\begin{eqnarray}
  \braket{Q}_{\mathrm{obs}} &=& -\frac{\partial \braket{E}_{\mathrm{obs}}}{\partial P},\\
  \braket{Q} &=& -\frac{\partial \Omega[T]}{\partial P} \approx
  \braket{Q}_{\mathrm{sim}}
\end{eqnarray}
where $\braket{Q}$ is the exact average of $Q(\mathbf{X})$ obtained from the partition function, and approximated by the simulation average $\braket{Q}_{\mathrm{sim}}$.
Based on these relations, we have the following procedure for parameter optimization.
\begin{enumerate}
  \item Initialize $J$ and $K$ with some values.
  \item Estimate $\braket{n_{YY'}^{b}}_{\mathrm{sim}}$ and $\braket{n_{Y}n_{Y'}}_{\mathrm{sim}}$ by running equilibrium MC simulations under the natural condition.
\item Update $P$ (either $J(Y,Y')$ or $K(Y,Y')$) by using the observed and simulated average values of $Q$ (respectively, $\braket{n_{YY'}^b}$ or $\braket{n_Yn_{Y'}}$)
  \begin{eqnarray}
    R^{(\nu+1)} &:=&  \alpha \left[
      \braket{Q}_{\mathrm{obs}} - \braket{Q}_{\mathrm{sim}}
      \right] + \beta R^{(\nu)},\label{eq:JKupdate1}\\
    P^{(\nu+1)} &:= & P^{(\nu)} + R^{(\nu+1)}\label{eq:JKupdate2}
  \end{eqnarray}
  where $\alpha$ and $\beta$ are small parameters, and $R^{(\nu)}$ is a momentum term introduced to accelerate convergence \cite{Qian1999}.
  \item Iterate 2 and 3.
\end{enumerate}
The parameter $\alpha$ was set to 0.3, 0.1 or 0.01 depending on optimization stages (see Results), and $\beta$ was set to 0.95.

\subsection{Data preparation}
The multiple sequence alignment (MSA) and profile hidden Markov model (HMM) of the SH3 domain (Pfam PF00018) were downloaded from the Pfam database \cite{Pfam} (version 30.0). The MSA was based on the ``representative proteomes'' of 75\% sequence identity cutoff. Based on the profile HMM, the length of the model was set to 48.

A representative crystal structure was chosen for each family from the Protein Data Bank \cite{wwPDB} based on the criteria that there are no gaps (insert or delete) within the domain and the resolution is better than 2.0 \AA.
This is done by first querying the PDBj Mine2 relational database \cite{KinjoETAL2017}, then by examining the alignments of the found PDB chains against the Pfam profile HMM using the HMMER hmmalign program \cite{HMMER3}. Following this procedure, I selected chicken Src SH3 domain (PDB: 4HVU \cite{PDB4HVU}) as a representative structure of the SH3 domain family, based on which there were 104 interacting non-bonded pairs of core sites.

\section{Results}
\subsection{Parameter optimization}
\begin{figure*}
  \includegraphics[width=\textwidth]{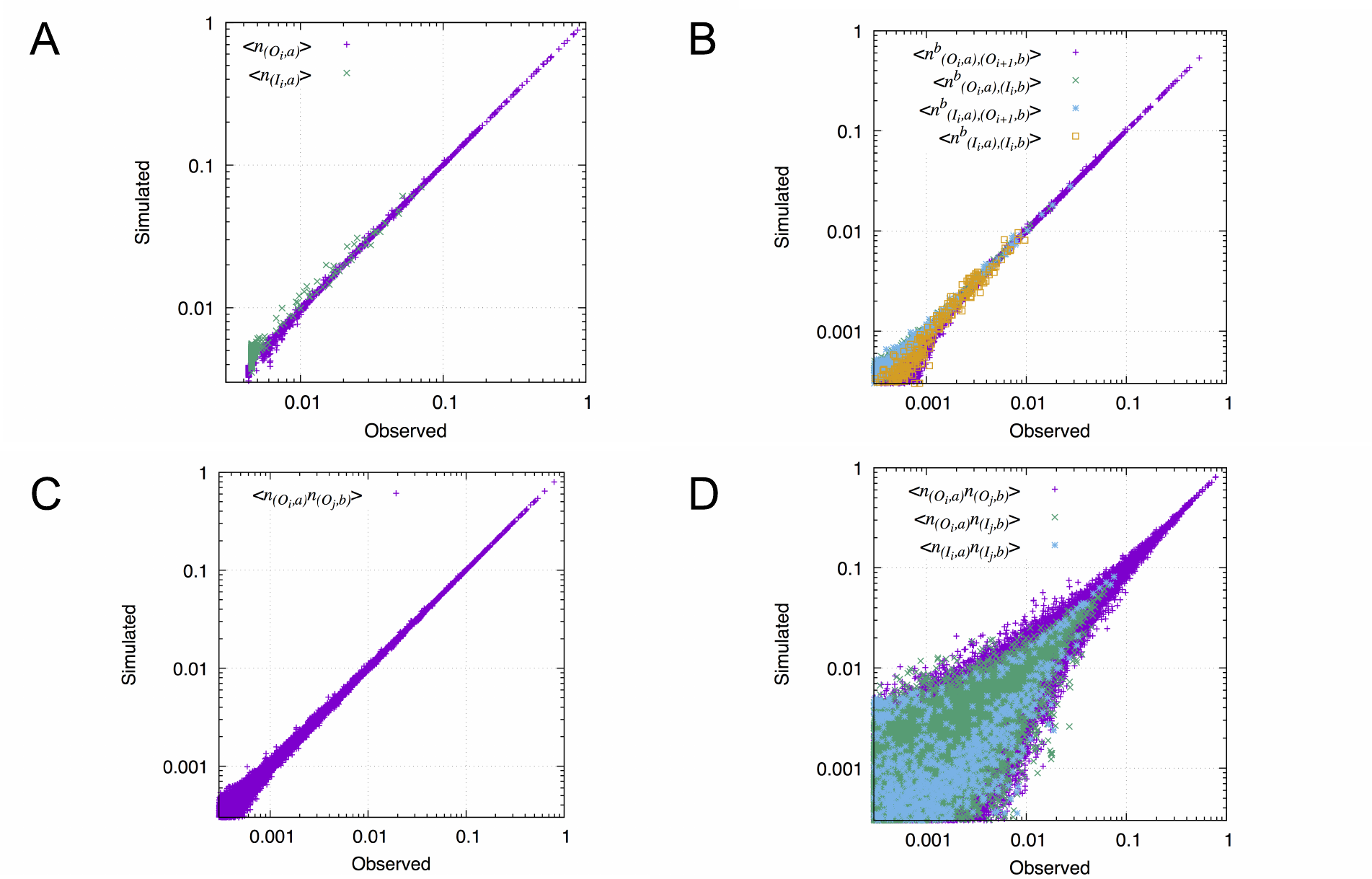}
  \caption{\label{fig:corr}Correlations between observed and simulated number densities.
    (A) Single-site number densities.
    (B) Bonded pair number densities.
    (C) Non-bonded pair number densities.
    (D) Non-bonded pair number densities for non-interacting pairs (these number densities were not optimized).}
\end{figure*}

I optimized the parameters $J$ and $K$ of the LGM model based on
the MSA of representative sequences of the SH3 family (Pfam PF00018) under
the natural condition ($T=1$ and $\mu(Y) = 0$ for all $Y \in \mathcal{Y}$).
The whole optimization process consisted of 3 stages. In the first stage, 120,000 sequences obtained from 12 trajectories of 10,000 sweeps each were used for computing the simulated densities, and then the parameters $J$ and $K$ were updated with $\alpha = 0.3$ (Eq. \ref{eq:JKupdate1}). This optimization was repeated for 500 steps. In the second stage, 1,200,000 sequences from 12 trajectories of 100,000 sweeps each were for used for updating the parameters with $\alpha = 0.1$. This stage consisted of 200 steps. The third stage was identical to the second stage except that $\alpha = 0.01$.

By construction, a sufficiently optimized LGM reproduces the number densities of each model site as well as of bonded and (interacting) non-bonded pairs in the given MSA. This is indeed demonstrated in Fig. \ref{fig:corr}. For the single-site number densities, the correlation coefficient between observed and simulated values were greater than 0.99 ($>$0.9999 for the core sites, and 0.993 for the insert sites). The correlation is slightly weaker for the insert sites, suggesting the difficulty of sampling the arbitrarily large number of inserted residues (there can be infinite number of inserted residues in theory, but the number is always limited in simulations). Note that the single-site densities were not directly optimized, but they were optimized indirectly through the bonded and non-bonded pair densities (c.f., Eqs. \ref{eq:relb1}-\ref{eq:relnb2}). For bonded pairs, the correlation coefficients were $>$ 0.999, 0.993, 0.994 and 0.984 for
$\braket{n_{(O_i,a),(O_{i+1},b)}^b}$,
$\braket{n_{(O_i,a),(I_{i},b)}^b}$, $\braket{n_{(I_i,a),(O_{i+1},b)}^b}$ and $\braket{n_{(I_i,a),(I_{i},b)}^b}$, respectively. For the interacting non-bonded pair densities, the correlation was $>$0.999. We use these parameters in the following.

For comparison, I also examined the correlation between non-interacting non-bonded pairs which were left out from the optimization (Fig. \ref{fig:corr}D). The simulated values do correlate with the observed values, but, as expected, the correlation is not very high: 0.991 for $\braket{n_{O_i}(a)n_{O_j}(b)}$, $\sim$0.82 for both $\braket{n_{O_i}(a)n_{I_j}(b)}$ and $\braket{n_{I_i}(a)n_{I_j}(b)}$. The pairs involving insert sites are less correlated, indicating less sufficient sampling for insert sites.

\subsection{Grand canonical MC simulations at various temperatures}
\begin{figure*}
  \includegraphics[width=\textwidth]{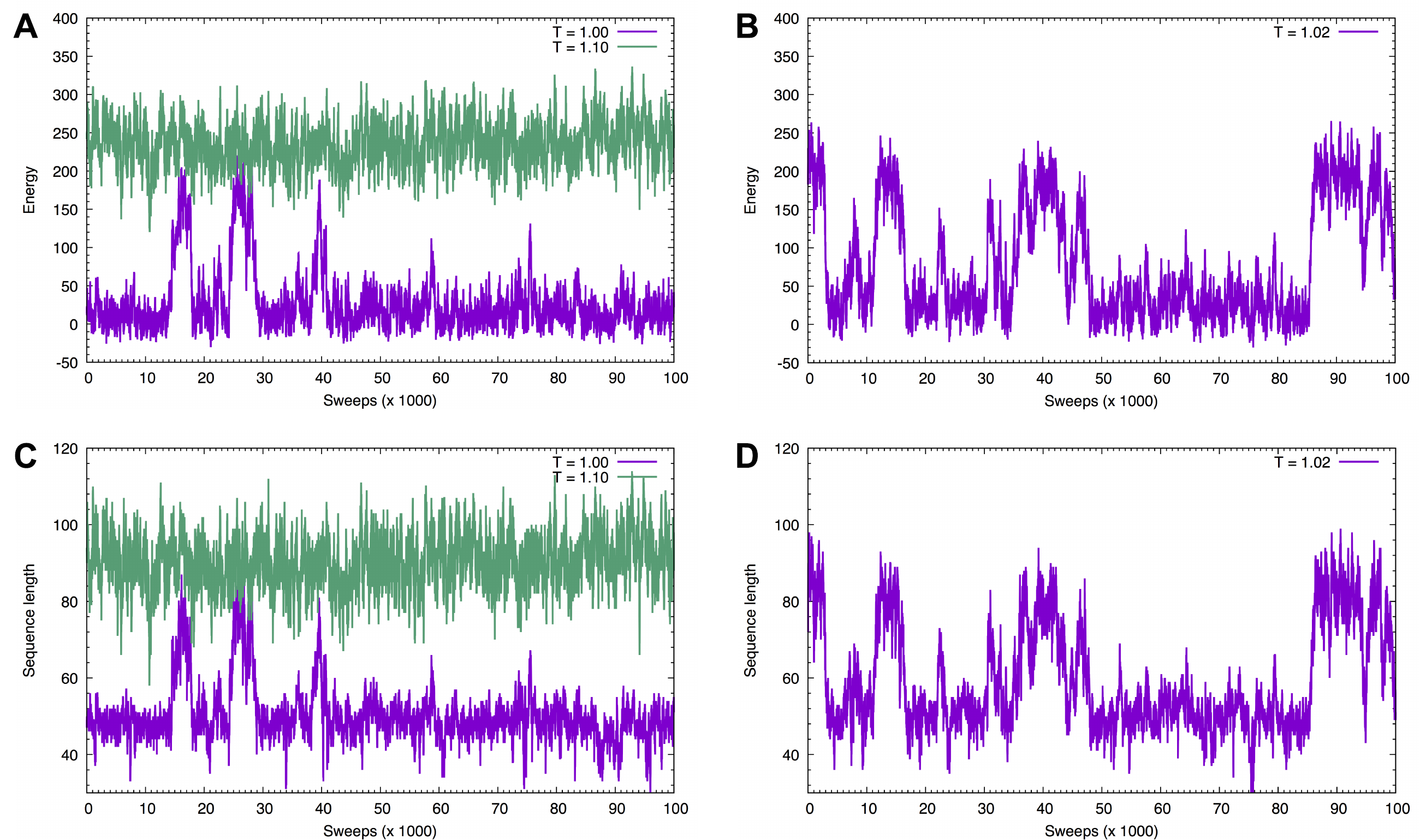}
  \caption{\label{fig:mctrajT}Trajectories of MC simulations under the standard condition ($\mu(Y) = 0$ for all $Y$).
    (A) Trajectories of energy from simulations at $T=1$ (magenta) and $T=1.1$ (green).
    (B) A trajectory of energy from a simulation at $T = 1.02$.
    (C) Trajectories of sequence length from the same simulations as in (A).
    (D) A trajectory of sequence length from the same simulation as in (B).
  }
\end{figure*}
Using the parameters defined above, MC simulations were performed at 3 different temperatures under the standard condition (i.e., $\mu(Y) = 0$ for all aligned sites). At $T = 1$ (the natural condition), the energy fluctuated mostly around 20-30 energy units (e.u.) with average 27.6 and standard deviation (s.d.) 6.4. The energy occasionally jumped to around 200 e.u., but very soon returned to the lower region (Fig. \ref{fig:mctrajT}A, magenta line). At a high temperature ($T = 1.1$), the energy steadily fluctuated around the average value of 239.2 e.u. with s.d. of 5.3 (Fig. \ref{fig:mctrajT}A, green line). At an intermediate temperature of $T = 1.02$, a two-state transition was observed (Fig. \ref{fig:mctrajT}B) between low-energy and high-energy states.

One of the strengths of the LGM is that it can handle variable-length insertions. We define the sequence length of an alignment as the number of standard residues in the alignment (i.e., the total number of non-delete ``residues''). The trajectories of sequence length show a large variety depending on the temperature (Figs. \ref{fig:mctrajT}C,D). At $T = 1$ and 1.1, the average sequence lengths were 48.8 (s.d. 4.6) and 91.5 (s.d. 6.9), respectively (Fig. \ref{fig:mctrajT}C).
The fluctuation of the sequence length is clearly correlated with that of the energy (Figs. \ref{fig:mctrajT}A,C). A similar trend is also observed for $T = 1.02$ (Fig. \ref{fig:mctrajT}D).

To characterize the sequences generated at different temperatures, I performed homology searches using the hmmsearch program \cite{HMMER3} using the profile HMM of the SH3 domain (SH3\_1.hmm provided by the Pfam database) against a database of sequences generated at a specified temperature. At $T = 1.00$, 8,961 out of 10,000 (89.6\%) of the sequences were significantly similar to the SH3 domain (E-value less than 0.01) whereas at $T = 1.1$, no sequence was significantly similar. This confirms that not only the residue distribution on average, but also the MC-generated sequences at the natural conditions are similar to the natural sequences. At $T=1.02$, the fraction of significantly similar sequences was 41.3\%. Thus, the transition observed in Fig. \ref{fig:mctrajT}B indeed indicates an order-disorder transition in the sequence space. In the following, ``ordered'' or ``disordered'' sequences refer to those similar or dissimilar, respectively, to the natural sequences. Since the ordered sequences are likely to fold into the native fold of the SH3 domain whereas the disordered sequences are not, the transition may be regarded as a ``folding'' transition in the sequence space.

\subsection{Perturbation on insert sites}
\begin{figure*}
  \includegraphics[width=\textwidth]{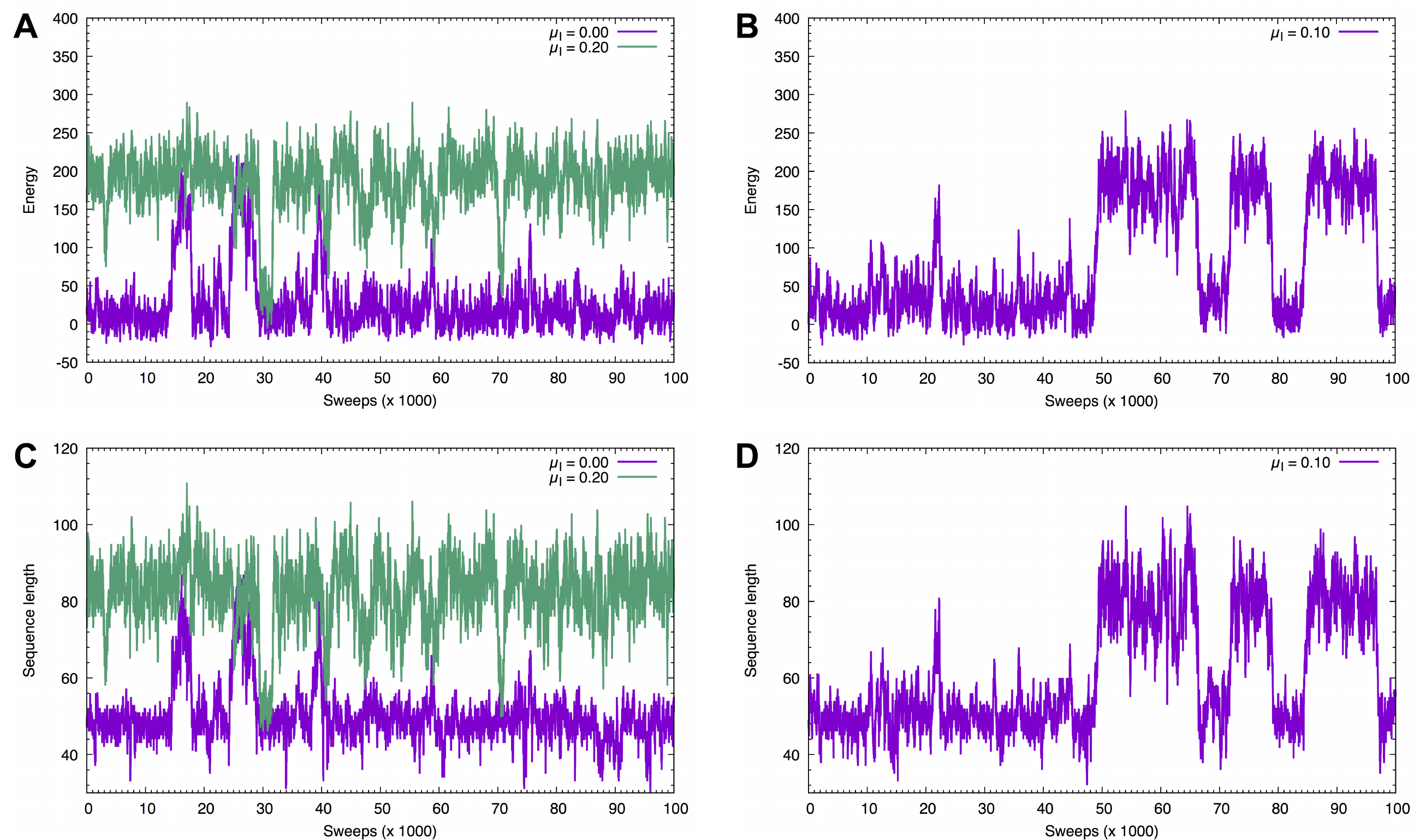}
  \caption{\label{fig:mctrajI}Trajectories of MC simulations with different $\mu_I$ values at $T = 1$.
    (A) Trajectories of energy from simulations at $\mu_I = 0$ (magenta) and $\mu_I=0.2$ (green).
    (B) A trajectory of energy from a simulation at $\mu_I = 0.1$.
    (C) Trajectories of sequence length from the same simulations as in (A).
    (D) A trajectory of sequence length from the same simulation as in (B).
  }
\end{figure*}

In addition to changing the temperature, we can also perturb the system
by introducing non-zero values for some or all of chemical potentials $\mu(Y)$.
Here, I show the results of simulations with $\mu(I_i,a) = \mu_I$, that is,
the chemical potential was set to a constant value $\mu_I$ for all residue types $a$ for all insert sites $I_i$ with $T = 1$ (Fig. \ref{fig:mctrajI}).
The case $\mu_I = 0$ (with $T = 1$) is the natural condition and the trajectory is identical to the one shown in Figs. \ref{fig:mctrajT}A,C (magenta lines).
When $\mu_I$ is set to a large value $\mu_I = 0.2$, the energy (Fig. \ref{fig:mctrajI}A) and the sequence length (Fig. \ref{fig:mctrajI}C) mostly had large values as expected. At an intermediate value $\mu_I = 0.1$, the trajectories again exhibited two-state transitions (Figs. \ref{fig:mctrajI}B,D).

Out of the 10,000 sequences generated with $\mu_I = 0.2$ and $\mu_I = 0.1$, 2.5\% and 46\%, respectively, were significantly similar to the profile HMM of the SH3 domain. Since the chemical potentials at the core sites were kept to zero, this result indicates that prolonged insertions alone can trigger order-disorder transition.

\subsection{Virtual temperature-jump experiment}
\begin{figure*}
  \includegraphics[width=0.5\textwidth]{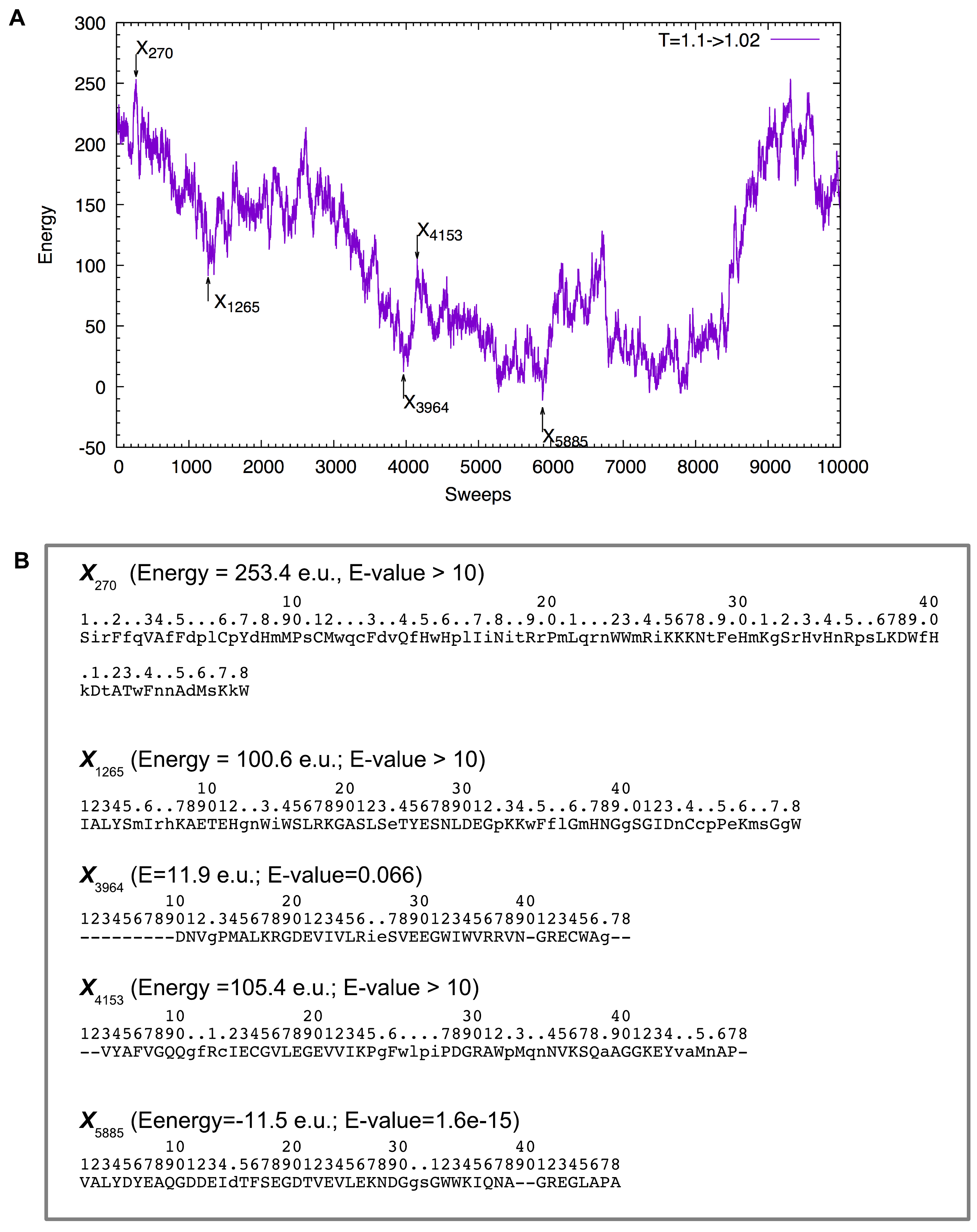}
  \caption{\label{fig:folding}A virtual temperature-jump experiment. The system was equilibrated at $T = 1.1$ for 50,000 sweeps, then the temperature was suddenly shifted to $T=1.02$. (A) The energy trajectory after the temperature jump.
    (B) 5 sequences at the steps marked in (A) are shown with  energy value of the sequence and the E-value for the alignment with the Pfam SH3 domain (estimated by the hmmsearch program \cite{HMMER3}). The residues in uppercase letters or ``-'' (delete) are those aligned with core sites (indexed by the core site position from 1 to 48, the length of the SH3 domain) and those in lowercase indicate inserted residues (``indexed'' with dots).}
\end{figure*}
Motivated by the existence of the ``folding'' transition, I next performed a virtual ``temperature-jump'' experiment in which the system was initially equilibrated at a high temperature ($T = 1.1$) and then it was suddenly cooled to a lower temperature ($T = 1.02$) and the trajectory was monitored (Fig. \ref{fig:folding}). As noted above (Fig. \ref{fig:mctrajT}A), the system is disordered (non-natural-like) at $T = 1.1$. An example, the alignment $\mathbf{X}_{270}$ (Fig. \ref{fig:folding}B), of the sequences shortly after the temperature jump illustrates some typical characteristics of high-energy, disordered sequences such as an excess of inserts and the lack of conserved residues (e.g., W32 and P47; c.f. $\mathbf{X}_{5885}$ below in Fig. \ref{fig:folding}B). As the ``folding'' proceeds, the sequence length tended to shrink ($\mathbf{X}_{1265}$ in Fig. \ref{fig:folding}B), and some conserved residues (e.g., W32) appeared when the energy became lower ($\mathbf{X}_{3964}$). However, the alignment still fluctuated to higher energies ($\mathbf{X}_{4153}$) before the entire region was ordered enough to significantly match the SH3 domain ($\mathbf{X}_{5885}$). Since $T=1.02$ is close to the transition temperature as suggested in the previous subsection (Fig. \ref{fig:mctrajT}B), the alignment continued to fluctuate largely.

\subsection{Mean-field approximation}
One of the motivations for developing the MC method for the LGM model was that the mean-field approximation did not work well in the previous study \cite{Kinjo2016} in which the parameters corresponding to $K$ in Eq. (\ref{eq:energy0}) were obtained by inverting the covariance matrix of the single-site counts of the core sites \cite{MorcosETAL2011}. This matrix inversion method has been derived from a mean-field approximation \cite{Plefka1982,Kinjo2015}. Thus, the parameters based on the mean-field approximation was not consistent with the mean-field approximation of the LGM model. Note that the approximations are involved in two different situations, the first one in determining the parameters and the second one in computing the partition function. A detailed account for various approximations in the first case has been provided by Cocco et al. \cite{CoccoETAL2017}
Since the parameters are determined rigorously by MC simulations (within the sampling and numerical errors) in the present study, we can examine if it is indeed the case that the mean-field approximation of the LGM model is inappropriate in the second case. I applied the mean-field approximation to calculate the partition function of the LGM model using the parameters obtained from MC simulations.
Please refer to the previous paper for the method to calculate the partition function using a transfer matrix method (with respective modifications to reflect the current model structure). Here, the mean-field of interactions is defined by
\begin{equation}
  \label{eq:mf}
  \tilde{K}_{(O_i,a)} = \sum_{(O_j,b)}K((O_i,a),(O_j,b))\braket{n_{(O_j,b)}}_{\mathrm{mf}}
\end{equation}
where the summation is over the partners of the interacting non-bonded pairs.
Note in particular that unlike the previous study \cite{Kinjo2016}, diagonal terms of the $K$ matrix are not included. The single-site number densities $\braket{n_{(O_j,b)}}_{\mathrm{mf}}$ are given by
\begin{equation}
  \label{eq:scf}
  \braket{n_{(O_j,b)}}_{\mathrm{mf}} = \sum_{\mathbf{X}}{n_{(O_j,b)}(\mathbf{X})P_{\mathrm{mf}}(\mathbf{X})}
\end{equation}
where the probability of alignment $P_{\mathrm{mf}}(\mathbf{X})$ (c.f., Eq. \ref{eq:boltzmann}) is computed using a transfer matrix method with the mean-field (Eq. \ref{eq:mf}) (see the previous paper \cite{Kinjo2016} for the details). Eqs. (\ref{eq:mf}) and (\ref{eq:scf}) are mutually dependent, and therefore, are solved mutually consistently.
For core sites, the difference of the mean-field or MC-generated single-site number densities from the observed ones was measured by the Kullback-Leibler (KL) divergence:
\begin{equation}
  \label{eq:kldiv}
  D_{O_i} = \sum_{a}\braket{n_{(O_i,a)}}\ln\left[\braket{n_{(O_i,a)}}/\braket{n_{(O_i,a)}}_{\mathrm{obs}}\right]
\end{equation}
where $\braket{n_{(O_i,a)}}$ indicates either $\braket{n_{(O_i,a)}}_{\mathrm{sim}}$ (obtained from MC simulations) or $\braket{n_{(O_i,a)}}_{\mathrm{mf}}$ (obtained from the mean-field approximation). 

\begin{figure}
  \includegraphics[width=1.0\textwidth]{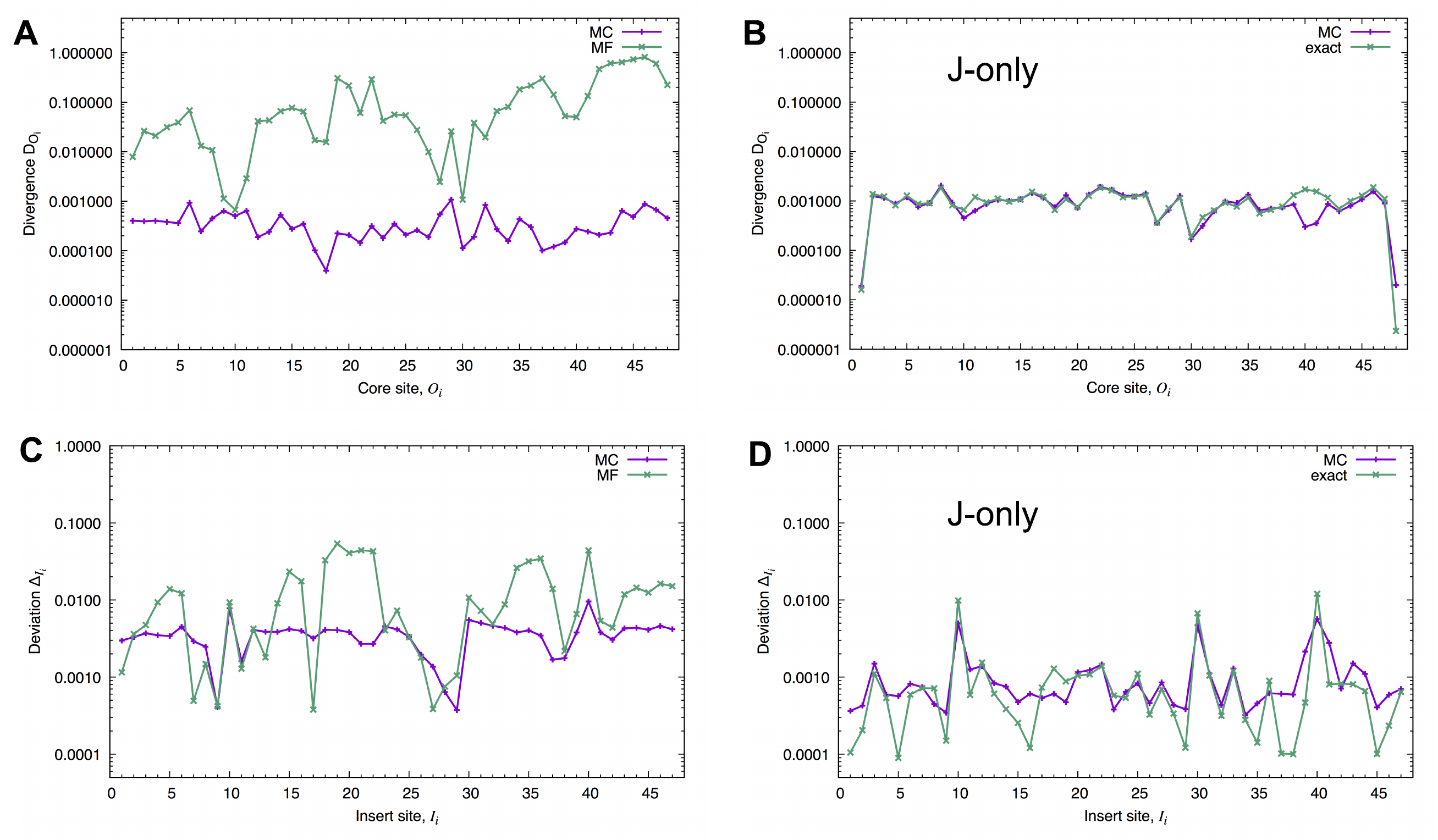}
  \caption{\label{fig:mf}Comparison between single-site number densities obtained from Monte Carlo simulations (MC) and the mean-field approximation (MF)
        as measured by the KL divergence from the observed distribution (Eq. \ref{eq:kldiv}) for the core sites (A and B) and root square deviation (Eq. \ref{eq:dev}) for the insert sites (C and D).
    Both MC and MF (and exact) results were obtained under the natural condition using the same parameters optimized by using MC simulations.
    (A) KL divergence of core sites for the ``full'' system with both $J$ and $K$ parameters optimized by MC simulations. (B) KL divergence of core sites for
    the ``$J$-only'' system where only the $J$ parameters were optimized and the long-range interactions between non-bonded pairs were ignored. In this case, the exact solution can be obtained \cite{Kinjo2016}.
    (C) Root square deviation of insert sites for the ``full'' system.
  (D) Root square deviation of insert sites for the ``$J$-only'' system. }
\end{figure}

The divergence of the core site number densities obtained from MC simulations  were less than $10^{-3}$ for most of the sites (the only exception was $1.07\times 10^{-3}$ for the core site 29) whereas those obtained from the mean-field approximation were greater than 0.01 for 41 out of 48 core sites (and $>0.1$ for 15 sites) and as large as 0.82 for the site 46 (Fig. \ref{fig:mf}A). To check this was not an artifact of the exact partition function, I also optimized the $J$ parameters without long-range interactions so the truly exact partition function could be obtained (Fig. \ref{fig:mf}B). For this ``$J$-only'' system \cite{Kinjo2016}, the simulation and exact results were very similar and the divergence values were all small, mostly less than $2\times10^{-3}$ (note that the
``exact'' result is not expected to yield  divergence of 0 due to the errors
in the parameters estimated from MC simulations).

For insert sites, we measure the difference between simulated and observed densities by the root square deviation instead of KL divergence (because the number densities are not normalized for insert sites):
\begin{equation}
  \label{eq:dev}
  \Delta_{I_i} = \sqrt{\sum_{a}(\braket{n_{(I_i,a)}}-\braket{n_{(I_i,a)}}_{\mathrm{obs}})^2}.
\end{equation}
The deviation of the insert site number densities obtained from MC simulations
were of order of $10^{-3}$, with the maximum value of $9\times 10^{-3}$ for $I_{40}$ whereas those obtained from the mean-field approximation were greater than 0.01 for 20 insert sites (Fig. \ref{fig:mf}C). For the ``$J$-only'' system, the deviations obtained from MC simulations and ``exact'' calculation were comparable (Fig. \ref{fig:mf}D). Since there are no non-bonded interactions involving insert sites, errors introduced to the core sites by the mean-field approximation also affect the insert sites through bonded interactions.

In summary, these results suggest that the mean-field approximation is indeed not appropriate for this model.

\section{Discussion}

In the present work, I used for long-range interactions only those involved in native 3D contacts. As such, the present model requires a priori knowledge of the native structure of the protein family of interest. While this may appear as a limitation in some respect (e.g., it cannot be used for structure prediction), it also has its advantages in other respects. First, parameter optimization is easier due to the smaller number of parameters. Since massive MC simulations are required to optimize a large number of  parameters, this is clearly an advantage \cite{LapedesETAL1999}.
Second, discarding non-bonded pairs not involved in native contacts eliminates statistical noise irrelevant to the structural context. This enables a simpler interpretation of the long-range interactions. Moreover, it becomes also possible to investigate the role of the native structure in determining the conservation patterns of the family sequences by comparing with systems optimized with ``misfolded'' structures. This may be an interesting subject for a future study.

I have shown that the mean-field approximation gives inconsistent results for the LGM model. It is, of course, expected that any approximations yield different results than the exact solution to some extent, and the difference may or may not be acceptable depending on the application. For studying the patterns of sequence conservation, however, the degree of inconsistency resulting from the mean-field approximation (Fig.\ref{fig:mf}A) is not acceptable. That is why in the previous study \cite{Kinjo2016} I resorted to the Gaussian approximation which includes self interaction (diagonal) terms in addition to long-range interactions. The self interaction terms are intrinsic to each site and residue, which should not exist in principle except for possibly integrated-out effects of intermolecular interactions such as ligand binding. In other words, each residue at a certain position does not have a means to ``know'' where in the sequence it is located and how it should be conserved other than by interactions with other residues at other positions (or ligands). Thus, it seems more appropriate to employ MC simulations for the present purpose. 

Using MC simulations, I have demonstrated the existence of a two-state transition in the sequence (sub)space around the SH3 domain. This sort of transition, analogous to the folding transition in the conformational space, has been suggested by Nishikawa \cite{Nishikawa1993,Nishikawa2002}. If this type of transition is universal for many protein families, it has some interesting theoretical as well as practical implications.
First, the boundary of each protein family can be determined clearly in terms of the transition point.
Note that most of the conventional sequence models such as profile HMMs are essentially one-dimensional models where sharp transitions are simply impossible \cite{Landau_Lifshitz_Stat_Phys}. Therefore, the boundary between family members and non-members is necessarily fuzzy according to the conventional models. This limitation of the conventional methods may have already biased our knowledge of protein families. 
Second, if the boundary of a protein family in the sequence space can be clearly defined, it may be possible to assign artificially designed protein sequences to existing protein families by following mutational paths along which the ordered state is maintained. This means that proteins belong to the same family whether they are artificially designed or naturally selected.
Third, it should be possible to characterize the transition state ensemble and to identify residues essential in determining protein families. Such characterization will be helpful in understanding evolutionary trajectories of proteins and what constitutes the protein family.

\section{Conclusion}
In summary, Monte Carlo simulations of the LGM model will be a convenient means to explore the structure of the protein sequence space. More thorough investigations on the ``folding'' transition in the sequence space using this model is under way.

\section*{Acknowledgments}
I thank Ikuo Fukuda and Ken Nishikawa for fruitful discussion.

\section*{Conflict of interest}
None declared.

\section*{Author contributions}
ARK did everything.

%\bibliography{refs,mypaper}

\end{document}